\documentclass{ptptex}

\usepackage{graphicx}

\title{Can a supernova bang twice?}

\author{J\"urgen~\textsc{Schaffner-Bielich}$^1$\footnote{invited talk 
given at the Yukawa International Program for Quark-Hadron Sciences: 
New Frontiers in QCD 2010, Kyoto, Japan},
Tobias~\textsc{Fischer}$^2$, 
Matthias~\textsc{Hempel}$^1$, 
Matthias~\textsc{Liebend\"orfer}$^3$,  
Giuseppe~\textsc{Pagliara}$^1$, 
Irina~\textsc{Sagert}$^4$}

\inst{$^1$Institut f\"ur Theoretische Physik, Philosophenweg 16,
  Ruprecht-Karls-Universit\"at, 69120 Heidelberg, Germany\\
  $^2$GSI Helmholtzzentrum f\"ur Schwerionenforschung, Planckstr. 1,\\
  64291 Darmstadt, Germany\\
  $^3$Physics Department, University of Basel, Klingelbergstr.~82,\\
  4056 Basel, Switzerland\\
  $^4$Institut f\"ur Theoretische Physik, Goethe-Universit\"at,
  Max-von-Laue-Str.~1, 60438~Frankfurt am Main, Germany}

\abst{The implications of a QCD phase transition at high temperatures
  and densities for core-collapse supernovae are discussed. For a
  strong first order phase transition to quark matter, various
  scenarios have been put forward in the literature. Here, detailed
  numerical simulations including neutrino transport are presented,
  where it is found that a second shock wave due to the QCD phase transition
  emerges shortly after bounce. It is demonstrated that such a
  supernova banging twice results in a second peak in the antineutrino
  spectrum. This second peak is clearly detectable in present neutrino
  detectors for a galactic supernova.}

\begin{document}

\maketitle

\section{Introduction}

Core-collapse supernova are enigmatic astrophysical cataclysmic
events. The observation of supernova SN1987A and the measurements of
its emitted neutrinos inspired to put forward unconventional explosion
mechanisms.  The detection of neutrinos was puzzling as there was a
discrepancy in the timing of the events from various neutrino
detectors. However, De Rujula pointed out that the measurements are
consistent with the emission of two separate neutrino bursts delayed
by about five hours.\cite{DeRujula:1987pg} He concluded that a
supernova may bang twice and suggested that a collapse to a black hole
triggered by accretion can explain the second neutrino peak. Hatsuda
argued that a hypothetical strange star, a quark star bound by
interactions only, could have formed within a second.\cite{Hatsuda:1987ck}
He pointed out that the released energy from
the transition to a strange star is comparable to the energy emitted
in supernovae. The phase transition to quark matter was studied for an
adiabatic collapse by Takahara and Sato.\cite{Takahara:1988yd} They
argued that the released latent heat increases the core temperature
drastically thereby generating a prolonged emission of neutrinos. In a
full hydro simulation with a phase transition a second shock wave
emerged \cite{Gentile:1993ma} but neutrinos were not considered in the
simulation. Drago and Tambini introduced another intriguing scenario
where the formation of strange quark matter leads to a prompt
bounce.\cite{Drago:1997tn} The mixed phase forms during the supernova
collapse, causes a softening of the equation of state. At higher
densities the equation of state stiffens again for being compatible
with the observed pulsar masses. Nakazato, Sumiyoshi, and Yamada
performed a full supernova simulation including neutrinos and the
quark-hadron mixed phase.\cite{Nakazato:2008su,Nakazato:2010ue}
They started with a
rather high initial mass of $100M_\odot$ finding no second shock wave
but only that the softening of the equation of state just shortened
the timescale for the collapse to a black hole. Here, we report on the
implementation of the QCD phase transition for core-collapse supernova
simulations with neutrinos included for lower initial progenitor
masses.\cite{Sagert:2008ka,Fischer:2010zz,Sagert:2010yg} A second
shock wave is generated by the appearance of the quark-hadron mixed
phase leading to an explosion. When the heated shock material is
running over the neutrino sphere, a second burst of neutrinos is
released in antineutrinos. This second peak can be observed with the
present neutrino detectors Super-K and IceCube for a galactic
supernova as shown by Dasgupta et al.~\cite{Dasgupta:2009yj}.


\section{QCD Phase Transition in Neutron Stars}


The QCD phase diagram is largely unknown at high densities and
temperatures.  The asymptotic freedom of QCD ensures that at large
enough energy scales matter should be described in terms of free
quarks and gluons and not by hadronic degrees of freedom.  At high
densities, quark matter could be in a chirally restored but not
deconfined phase which is dubbed the quarkyonic phase.\cite{Mclerran:2007qj}
Compact stars consisting of pure quark matter,
so called strange stars, could be bound just by interactions with
quite exotic properties as shown within the MIT bag model
\cite{Witten:1984rs,Haensel:1986qb,Alcock:1986hz} (see also
Ref.~\citen{Baym:1976yu}). In perturbative QCD calculations to ${\cal
  O}(\alpha_s^2)$
\cite{Freedman:1978gz,Fraga:2001id,Andersen:2002jz,Fraga:2004gz,Kurkela:2009gj}
the properties of these strange stars were found to be surprisingly
similar to the results of the MIT bag model. Strange stars have quite
similar maximum masses and radii compared to ordinary neutron
stars. However, a strange star is hypothetical as one has to assume
that strange quark matter is more stable than ordinary nuclear matter.

Hence, one has to match to hadronic matter at low densities. The onset
of the mixed phase could be as low as $(1-2)n_0$ for a large range of
values for the MIT bag constants, see e.g. Ref.~\citen{Schertler:2000xq}. Sufficiently high densities are reached in
the core for a $1.3M_\odot$ neutron star to have quark matter present.
For a strong first order phase transition a third family of compact
stars appears in the mass-radius diagram besides white dwarfs and
neutron stars.\cite{Glendenning:1998ag,Schertler:2000xq} Signals for
such a phase transition and the appearance of a third family have been
discussed intensively in the literature, as e.g.\ the spontaneous
spin-up of pulsars \cite{Glendenning:1997fy} and an exotic mass-radius
relation with the so called rising twins.\cite{Schertler:2000xq}
Also the collapse of a neutron star to the third family can release
gravitational waves, $\gamma$-rays, and neutrinos. The signal of the
QCD phase transition for core-collapse supernovae will be the subject
of the next section.


\section{QCD phase transition in core-collapse supernovae}


The conditions for phase equilibrium in supernova matter have been
devised in detail in Ref.~\citen{Hempel:2009vp}. Several different cases
have been considered depending on the locally and globally conserved
charges in the thermodynamic system at hand which for supernovae
matter would be the proton or lepton fraction in addition to the
Coulomb charge. An interesting case concerns the one, where the charge
is assumed to be locally conserved not globally. Then chargeless
bubbles appear in leptonized matter at the early stage of the
proto-neutron star's life which form a mixed phase due to a
nonvanishing locally conserved lepton number. The mixed phase
disappears when the neutrinos leave the star which can cause a delayed
collapse with a pronounced emission of neutrinos.\cite{Pagliara:2009dg}
Proto-neutron star evolution with quarks have been simulated in the
work of Pons et al.~\cite{Pons:2001ar}. They find that the onset of
the quark phase in core-collapse supernovae occurs during the late
time evolution of the proto-neutron star. The timescale for quark
matter to appear was found to be typically $(5-20)$s which is well
after the bounce. The late onset of the quark phase is due to a large
bag constant, $B^{1/4}>180$~MeV which was chosen to get a
large neutron star mass. 

There are several arguments for quark matter to appear at rather low
densities in astrophysical systems. First in $\beta$-equilibrium,
strange quark matter is favored in comparison to ordinary nuclear
matter due to the additional strangeness degree of freedom.  Secondly,
low values of the proton fraction are also an advantage for quark
matter due to the large asymmetry energy of nuclear matter (the
situation gets more subtle when effects of color-superconductivity are
considered, see Ref.~\citen{Pagliara:2010ii} for more details). Thirdly, the
phase transition line is located at lower densities at higher
temperatures. The total effect is that the production of quark matter
in supernovae material can occur at quite low densities so that its
production at bounce is possible. For a temperature of $T=20$~MeV the
phase transition to the mixed phase can be located just slightly above
normal nuclear matter density for $Y_p=0.3$. It is important to note
that for matter in heavy-ion collisions the phase boundary is shifted
to much larger values.  It is also crucial to realize that the phase
transition is not necessarily related to deconfinement, so the naive
picture of overlapping hadrons making the transition can be
misleading. In principle, any strong first-order phase transition can
be envisioned in dense matter for our purpose here, as one related to
chiral symmetry restoration.

There are only two supernova equations of state commonly used for the
nuclear phase, the one of Lattimer and Swesty \cite{Lattimer:1991nc}
and the one of Shen et al.~\cite{Shen:1998gq}. Advanced modern
equations of state are presently being developed. Medium effects for
nuclei have been studied in Ref.~\citen{Ropke:2008qk} and applied for
supernova conditions by Typel et al.~\cite{Typel:2009sy}.  The
statistical approach of Botvina and Mishustin for supernova matter
utilizes methods from nuclear multifragmentation.\cite{Botvina:2008su}
A complete supernova equation of state has been
developed with an excluded volume which takes into account the whole
set of nuclei of the nuclear chart in Ref.~\citen{Hempel:2009mc}. In the
following we will use the equation of state from Shen et al.\ where
the phase transition is added by using the MIT bag model.

\begin{figure}
\centerline{\includegraphics[width=0.5\textwidth]{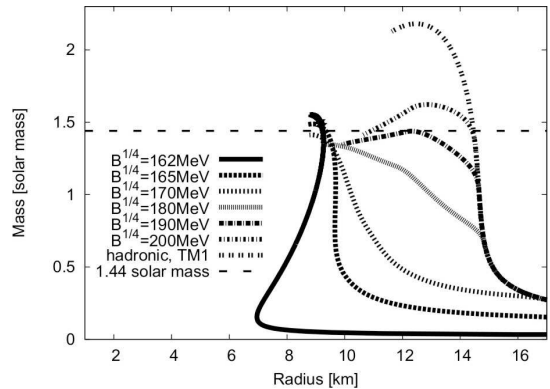}}
\caption{The mass-radius relation for cold hybrid stars for different
  values of the MIT bag constant and $\alpha_s$ corrections. Present
  constraints from pulsar mass measurements are for PSR1913+16 $1.44
  M_\odot$ and for PSR 1903+0327 $1.67 M_\odot$
  \cite{Champion:2008ge,Freire:2009fn} (figure taken
  from Ref.~\citen{Fischer:2010zz}).}
\label{fig:massradius}
\end{figure}

It turned out that a particular critical point to check the equation
of state is the maximum mass for cold neutron stars. Recent mass
measurements of the pulsar PSR 1903+0327 indicate a quite high mass of
$M=(1.67\pm 0.01)M_\odot$.\cite{Champion:2008ge,Freire:2009fn}
A detailed discussion on mass and radius constraints for cold neutron
stars can be found in Refs.~\citen{Lattimer:2004pg,Lattimer:2006xb,Steiner:2010fz}.
The mass-radius curve for the equation of state used in the supernova
simulation is depicted in Fig.~\ref{fig:massradius} for different
values of the MIT bag constant. A lower value results in a higher
maximum mass. Here we find maximum masses of $M_{\rm max}=1.56
M_\odot$ for $B^{1/4}=162$~MeV and $M_{\rm max}=1.5 M_\odot$ for
$B^{1/4}=165$~MeV.  If one includes corrections from one-gluon
exchange, the maximum mass increases to $M_{\rm max}=1.67 M_\odot$
(for $\alpha_s=0.3$) while the critical density for the onset of the
mixed phase is tuned to be similar by lowering the bag constant to
$B^{1/4}=155$~MeV.\cite{Fischer:2010zz,Sagert:2010yg}

The strong first order QCD phase transition has a significant impact
on the dynamical evolution of core-collapse supernovae. Full
hydrodynamical simulations including neutrinos have been performed for
progenitor masses of $M=10M_\odot$ and $M=15M_\odot$ in Ref.~\citen{Sagert:2008ka}.
A few hundred milliseconds after the bounce, a
second shock wave develops due to the formation of the new
high-density phase. The mixed phase has a lower adiabatic index,
collapses and hits the high-density core which has a much larger
adiabatic index.  A shock front forms, travels outwards and
accelerates at the steep density gradient on the surface of the
proto-neutron stars. When the heated material passes the
neutrinosphere, antineutrinos are released in a burst. The
antineutrino spectrum will be markedly different from the conventional
picture as a pronounced second peak appears shortly after the time of
collapse. In the simulations runs the quark core appears between
$t_{\rm pb}=200$ to $500$~ms depending on the progenitor mass and the
chosen bag constant.
The transition to strange quark matter involves the production of
strangeness during the conversion process. In hadronic matter,
hyperons can appear to some extent in supernova material due to
$\beta$-equilibrium and thermal production,
see Refs.~\citen{Ishizuka:2008gr,Sumiyoshi:2008kw} who find a hyperon fraction at
bounce (assuming $T\sim 20$~MeV) of about 0.1\%.  This small amount of
hyperons already present enables the nucleation of strange quark
matter via fluctuations of strangeness.\cite{Mintz:2009ay,Mintz:2010mh}
The results are highly sensitive to
the surface tension between the old and the new phase, which is a
basically unknown quantity. The nucleation timescales computed are
below the timescale of a supernova for critical surface tensions of
$\sigma < 20$ MeV~fm$^{-2}$, which are usually considered to be small
but which are not unreasonable.

\begin{figure}
\centerline{\includegraphics[width=0.34\textwidth]{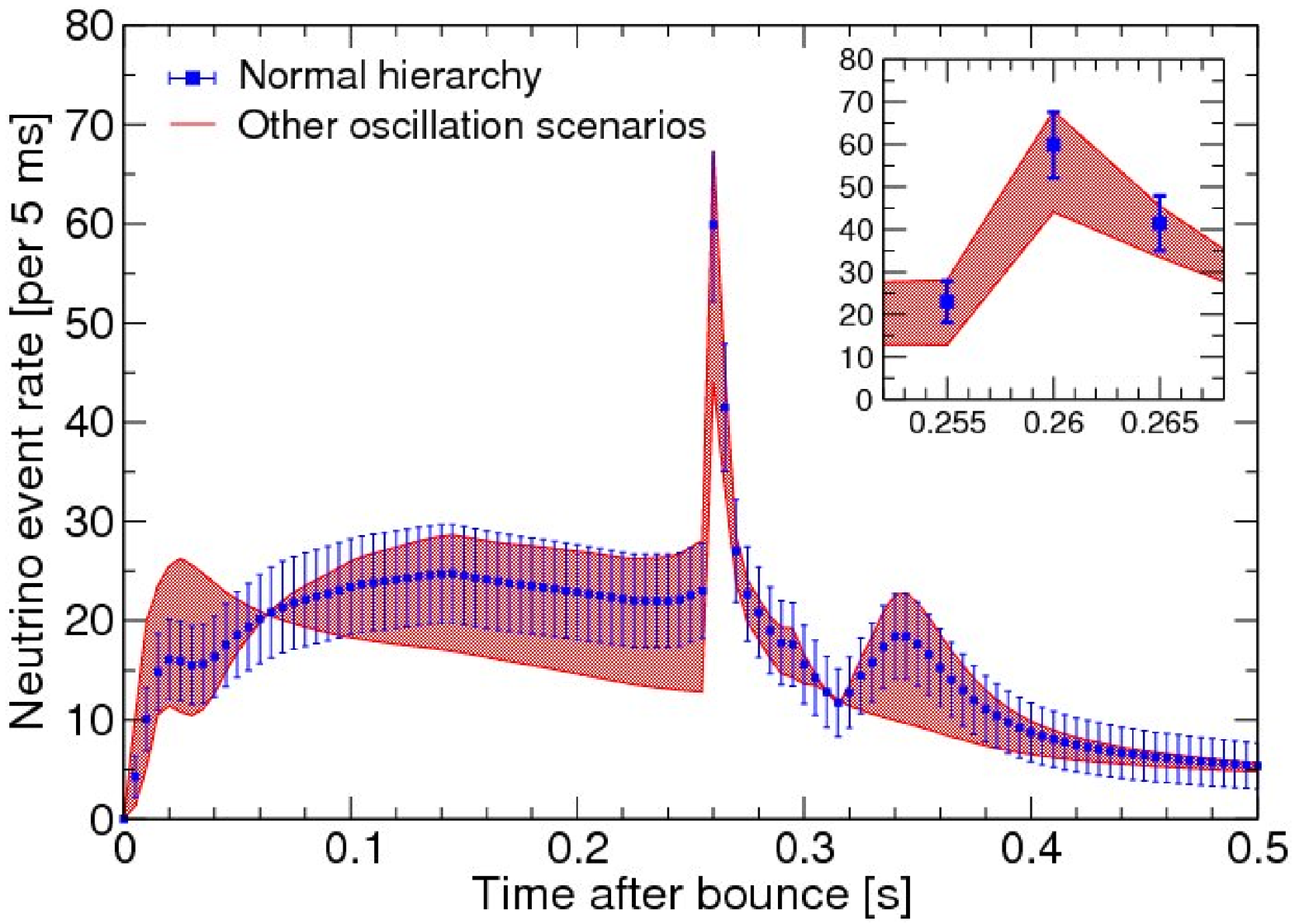} \qquad
\includegraphics[width=0.34\textwidth]{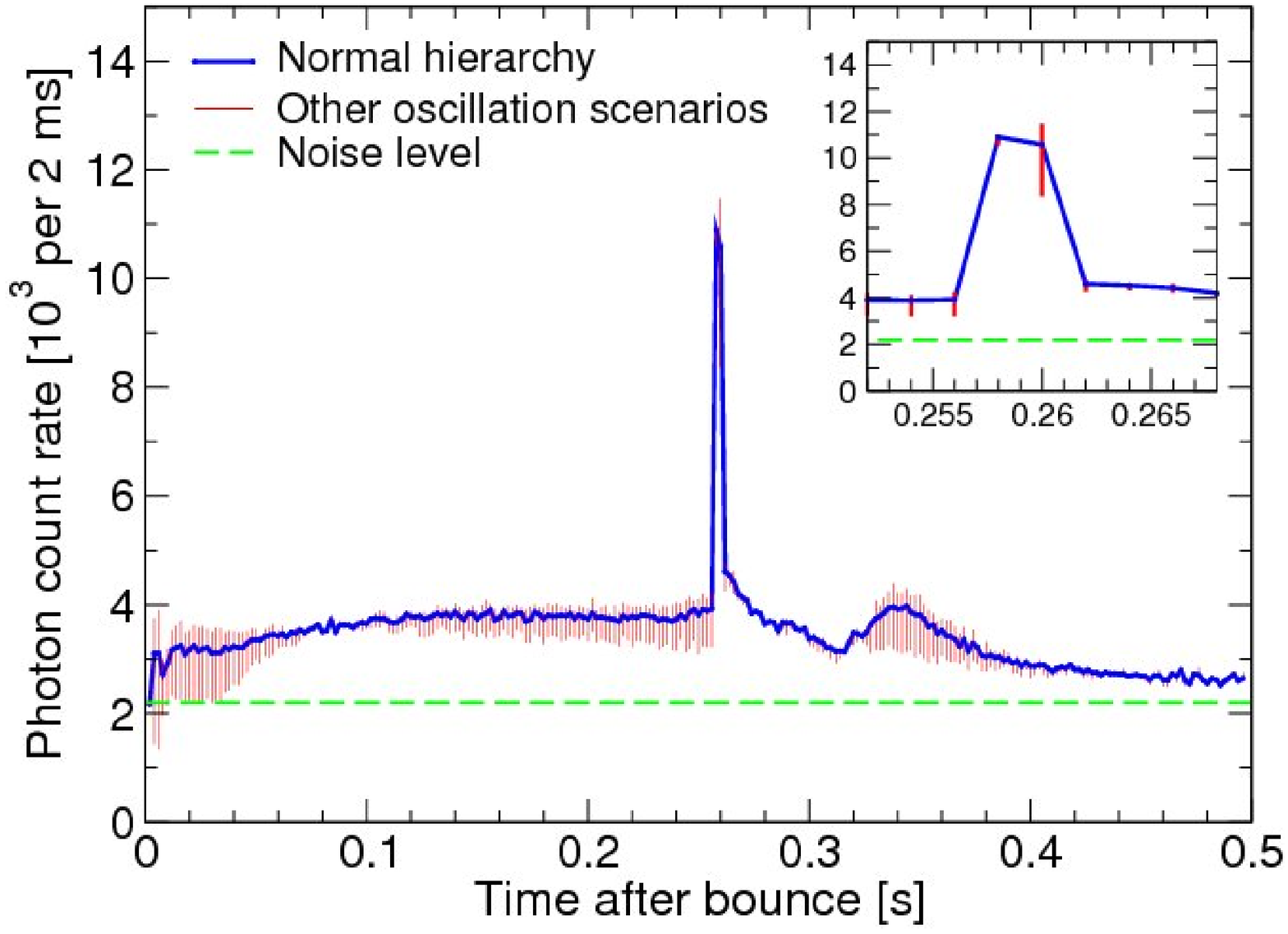}}
\caption{The expected signal of antineutrinos for a supernova banging
  twice at a galactic distance of 10~kpc as seen in the
  Super-Kamiokande detector (left plot) and in IceCube (right plot). A
  spectacular pronounced second peak is clearly seen in the simulated
  data, here at a time of around 260~ms (figures taken from Dasgupta
  et al.~\cite{Dasgupta:2009yj}).}
\label{fig:neutrinos}
\end{figure}

The antineutrino peak from the QCD phase transition can be detected
with the present neutrino detectors Super-K and IceCube, see Dasgupta
et al.~\cite{Dasgupta:2009yj}. These detectors are mostly sensitive to
antineutrinos by inverse $\beta$-decay reactions ($\bar \nu_e p \to n
e^+$). Adopting the antineutrino spectrum from the supernova
simulation, the observed antineutrino signal from a supernova would
clearly reveal a second peak. Fig.~\ref{fig:neutrinos} shows the
expected signal for Super-K and IceCube assuming a galactic supernova
at a distance of 10 kpc. The second peak in the antineutrino spectra
sticks out quite strikingly showing the high sensitivity of the
present neutrino detectors. Unfortunately, the sensitivity of the
neutrino detectors at the time of SN1987A were orders of magnitude
lower, otherwise one might have already seen that a supernova can
indeed bang twice.


\section*{Acknowledgements}

This work is supported by BMBF under grant FKZ 06HD9127, by DFG under
grant PA1780/2-1 and within the framework of the excellence initiative
through the Heidelberg Graduate School of Fundamental Physics, the
Gesellschaft f\"ur Schwerionenforschung GSI Darmstadt, the Helmholtz
R
esearch School for Quark Matter Studies, the Helmholtz Graduate
School for Heavy-Ion Research (HGS-HIRe), the Graduate Program for
Hadron and Ion Research (GP-HIR), the Helmholtz Alliance Program of
the Helmholtz Association, contract HA-216 "Extremes of Density and
Temperature: Cosmic Matter in the Laboratory", the Swiss National
Science Foundation, grant no.~PP00P2-124879/1 and 200020-122287, and
CompStar, a research networking program of the European Science
Foundation.


\bibliographystyle{utphys}
\bibliography{all,literat}

\end{document}